\documentclass[aps,preprint]{revtex4}%
\usepackage{amsfonts}
\usepackage{amsmath}
\usepackage{amssymb}
\usepackage{graphicx}%
\providecommand{\U}[1]{\protect\rule{.1in}{.1in}}

\begin{document}
\title{Two-dimensional $\chi^{2}$ solitons generated by the downconversion of Airy waves}
\author{Thawatchai Mayteevarunyoo}
\affiliation{Department of Telecommunication Engineering, Mahanakorn University of
Technology, Bangkok 10530}
\author{Boris A. Malomed}
\affiliation{Department of Physical Electronics, School of Electrical Engineering, Faculty
of Engineering, Tel Aviv University, Tel Aviv 69978, Israel}
\pacs{PACS number}

\begin{abstract}
Conversion of truncated Airy waves (AWs) carried by the second-harmonic (SH)
component into axisymmetric $\chi^{2}$ solitons is considered in the 2D system
with the quadratic nonlinearity. The spontaneous conversion is driven by the
parametric instability of the SH wave. The input in the form of the AW vortex
is considered too. As a result, one, two, or three stable solitons emerge in a
well-defined form, unlike the recently studied 1D setting, where the picture
is obscured by radiation jets. Shares of the total power captured by the
emerging solitons and conversion efficiency are found as functions of
parameters of the AW input.

\end{abstract}
\maketitle

Intrinsic structure of waves may steer their propagation along curved
trajectories, a well-known example being Airy waves (AWs) in linear media
\cite{Berry}-\cite{Chr2}. Realizations of AWs have been predicted and
demonstrated in optics \cite{Chr1}-\cite{Milivoj}, plasmonics \cite{plasm1}-%
\cite{plasm3}, electron beams \cite{el}, 
and gas discharge \cite{discharge}. In linear media, the propagation of
two-dimensional (2D) factorized AWs has been demonstrated too \cite%
{Berry,Chr1,Chr2,1Dand2Dexperiment0,1Dand2Dexperiment}.

Considerable work has been done on the dynamics of AWs in media with cubic
\cite{ChrSegev-nonlin}-\cite{Radik} and quadratic ($\chi ^{(2)}$) \cite%
{Ady-3wave}-\cite{Ady-Moti} nonlinearities, including generation of solitons
by AWs, and the 2D propagation in a $\chi ^{(2)}$ medium \cite%
{2DexperimentChi2}. Most works dealing with the $\chi ^{(2)}$ nonlinearity
addressed the \textit{upconversion} scenario, i.e., the AW was launched in
the fundamental-frequency (FF) component, generating the second-harmonic
(SH) wave \cite{rev1}-\cite{rev3}. The \textit{downconversion} of the 1D
truncated AW launched in the SH, which spontaneously generates the FF
component due to the parametric instability of the SH wave, was recently
considered in Ref. \cite{we}. The spontaneous downconversion, initiated by
random perturbations amplified by the parametric instability \cite{down1D,down2D},
produces sets of one, two, or three solitons alternating with iregular
radiation \textquotedblleft jets". The generation of the solitons from the
SH AW establishes the direct dynamical link between two distinct types of
\textit{eigenmodes} in the quadratic medium, \textit{viz}., the single-color
AWs and two-color solitons.

The objective of the present work, suggested by the stability of fundamental
2D $\chi ^{(2)}$ solitons \cite{rev1}-\cite{rev3} (the first reported $\chi
^{(2)}$ solitons were two-dimensional \cite{2Dchi2solitons}), is to
establish such a link in 2D, via the spontaneous downconversion of
two-dimensional AWs. The results are essentially \textquotedblleft cleaner"
than in the 1D case, i.e., emerging solitons are well pronounced, with no
conspicuous radiation features or irregularities obscuring the picture. We
also consider the downconversion of the input in the form of an AW vortex,
which is another single-color eigenmode in 2D. In accordance with the fact
that $\chi ^{(2)}$ vortex solitons are unstable \cite{vort1}-\cite{vort3},
it is found that such an input generates solely fundamental solitons.

The 2D spatial-domain evolution of the FF and SH wave amplitudes, $u$ and $w$%
, is governed by well-known scaled equations \cite{rev1}-\cite{rev3}:%
\begin{gather}
iu_{z}+(1/2)\nabla ^{2}u+u^{\ast }w=0,  \label{u} \\
2iw_{z}-qw+(1/2)\nabla ^{2}w+u^{2}/2=0,  \label{w}
\end{gather}%
where $z$ is the propagation distance, $(x,y)$ transverse coordinates, $%
\nabla $ the respective gradient, and the mismatch parameter may be scaled
to three values: $q=-1,0,+1$. The system conserves the total power,
vectorial momentum, and angular momentum (as well as the Hamiltonian):%
\begin{equation}
P=\int \int \left( |u|^{2}+4|w|^{2}\right) dxdy,  \label{P}
\end{equation}%
\begin{equation}
{\bf M}=i\int \int \left[ \left( u\nabla u^{\ast
}+2w\nabla w^{\ast }\right) -\mathrm{c.c.}\right] dxdy,
\label{M}
\end{equation}
\begin{gather}
\Omega =\frac{i}{2}\int \int \left\{ \left[ u\left( x\frac{\partial }{%
\partial y}-y\frac{\partial }{\partial x}\right) u^{\ast }\right. \right.
\nonumber \\
\left. \left. +2w\left( x\frac{\partial }{\partial y}-y\frac{\partial }{%
\partial x}\right) w^{\ast }\right] -\mathrm{c.c.}\right\} dxdy,
\label{Omega}
\end{gather}%
where $\mathrm{c.c.}$ stands for the complex-conjugate expression.

In the absence of the FF component, exact 2D eigenmodes are generated by the
SH input in the form of factorized truncated AWs,

\begin{gather}
w\left( x,y,z=0\right) =W_{0}\mathrm{Ai}\left( \alpha x\right) \mathrm{Ai}%
\left( \alpha y\right) e^{\left( \alpha /A\right) \left( x+y\right) }
\nonumber \\
\equiv \mathrm{AA}\left( x,y\right) ,~v(x,y,z=0)=0.  \label{initial}
\end{gather}%
where $\mathrm{Ai}$ is the Airy function, with $W_{0}$ and $1/\alpha $
determining the amplitude and spatial scale of the wave, whose total power
is made finite by means of the truncation parameter, $A$ \cite{Chr1}:%
\begin{equation}
P=W_{0}^{2}\left( 8\pi \alpha ^{2}\right) ^{-1}A\exp \left[ \left(
4/3\right) A^{-3}\right] .  \label{Power}
\end{equation}%
More general solutions may be produced by anisotropic inputs, with $\mathrm{%
Ai}\left( \alpha y\right) $ in Eq. (\ref{initial}) replaced by $\mathrm{Ai}%
\left( \beta y\right) $, $\beta \neq \alpha $.

Fundamental solutions of Eqs. (\ref{u}) and (\ref{w}) are 2D isotropic
solitons, whose existence and stability for large $q>0$ can be easily
explained by means of the cascading limit \cite{rev1}: in this case, a
solution to Eq. (\ref{w}) is approximated, at the first two orders, by $w=%
\left[ (2q)^{-1}+\left( 2q\right) ^{-2}\nabla ^{2}\right] (u^{2}).$ The
substitution of this in Eq. (\ref{u}) leads to a generalized nonlinear Schr%
\"{o}dinger (NLS) equation,
\begin{equation}
iu_{z}+(1/2)\nabla ^{2}u+\left( 2q\right)
^{-1}|u|^{2}u+\left( 2q\right) ^{-2}u^{\ast }\nabla ^{2}\left(
u^{2}\right) =0.
\label{NLS}
\end{equation}
The last term here, which corresponds to term $\delta \mathcal{H%
}=\left( 8q^{2}\right) ^{-1}\left\vert \nabla (u^{2})\right\vert ^{2}$ in
the Hamiltonian density, is similar to, but different from, the one which
accounts for a finite interaction radius in the Gross-Pitaevskii equation
for atomic Bose-Einstein condensates, \textit{viz}., $u\nabla ^{2}(|u|^{2})$
\cite{Spain}). A usual estimate \cite{Fibich} demonstrates that this term
stabilizes 2D NLS solitons by arresting their collapse. It is shown below
that the AW structure of the input, driving self-stretching of the field,
prevents approaching the quasi-collapse in the $\chi^{(2)}$ system,
and thus facilitates the creation of one or several solitons
(a similar trend was reported in the cubic NLS equation \cite{Radik}).

Results produced by systematic simulations of Eqs. (\ref{u}) and (\ref{w})
with initial conditions given by Eq. (\ref{initial}) with the FF component
seeded by a small random perturbation with amplitude $\sim 0.001$, are
presented below for $q=0$ and $q=+1$. The outcome does not depend on a
particular realization of the random seed, which is explained by the fact
that the system selects an eigenmode of the parametric instability from
the random perturbation. In agreement with Eq. (\ref{NLS}),
solitons were not found at $q=-1$. The results are reported for $A\leq 20$, the
respective FWHM width of the input in the $x$ and $y$ directions being $\leq
30$; otherwise, the necessary integration domain is too large, and the
respective spatial area of the experimental realization may also be too
large.

\begin{figure}[t]
\centering\includegraphics[width=3in]{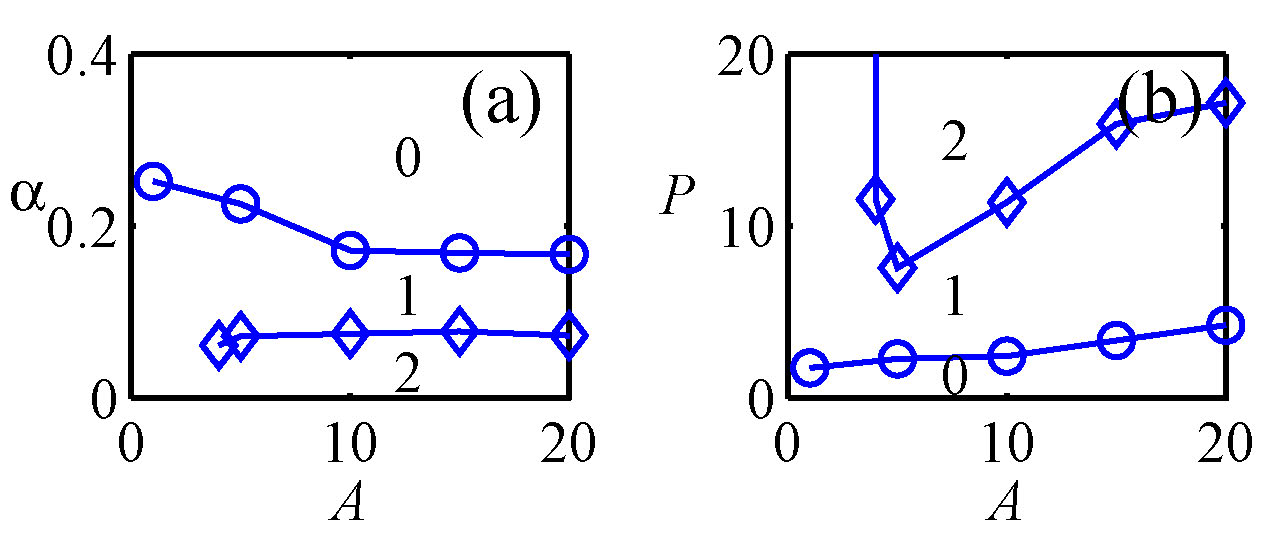}
\caption{(a,b): Projections of a typical slice of the diagram in the
parameter space $\left( A,\protect\alpha ,P\right) $ onto planes ($A,\protect%
\alpha $) and ($A,P$), showing the number of solitons ($0$, $1$, or $2$)
generated by input (\protect\ref{initial}) through the downconversion, in
the system with $q=0$.}
\label{Fig1}
\end{figure}

The results for $q=0$ are summarized in Fig. \ref{Fig1}, which shows the
number of solitons (none, one, or two) generated by the AW input, in the
space of its parameters, $\left( A,\alpha ,P\right) $. The full diagram in
the 3D parameter space looking cumbersome, Figs. \ref{Fig1}(a,b) display its
typical 2D slice projected onto the $\left( A,\alpha \right) $ and $\left(
A,P\right) $ planes. Naturally, the number of the solitons increases with $P$
in Fig. \ref{Fig1}(b). At $A\geq 10$, the results are simple in terms of $%
\alpha $: two, one, and no solitons are generated, respectively, at $\alpha
<0.075$, $0.075<\alpha <0.17$, and $\alpha >0.17$ in Fig. \ref{Fig1}(a).
Three solitons can be generated at very small $\alpha $ (this case is not
shown in Fig. \ref{Fig1}). In particular, three solitons with nearly equal
powers, each carrying share $\simeq 0.12$ of the total input power (\ref%
{Power}), cf. Fig. \ref{Fig5} below, are generated by the input with $%
W_{0}=0.39$, $\alpha =0.05$, $A=5$.

\begin{figure}[tb]
\centering\includegraphics[width=3in]{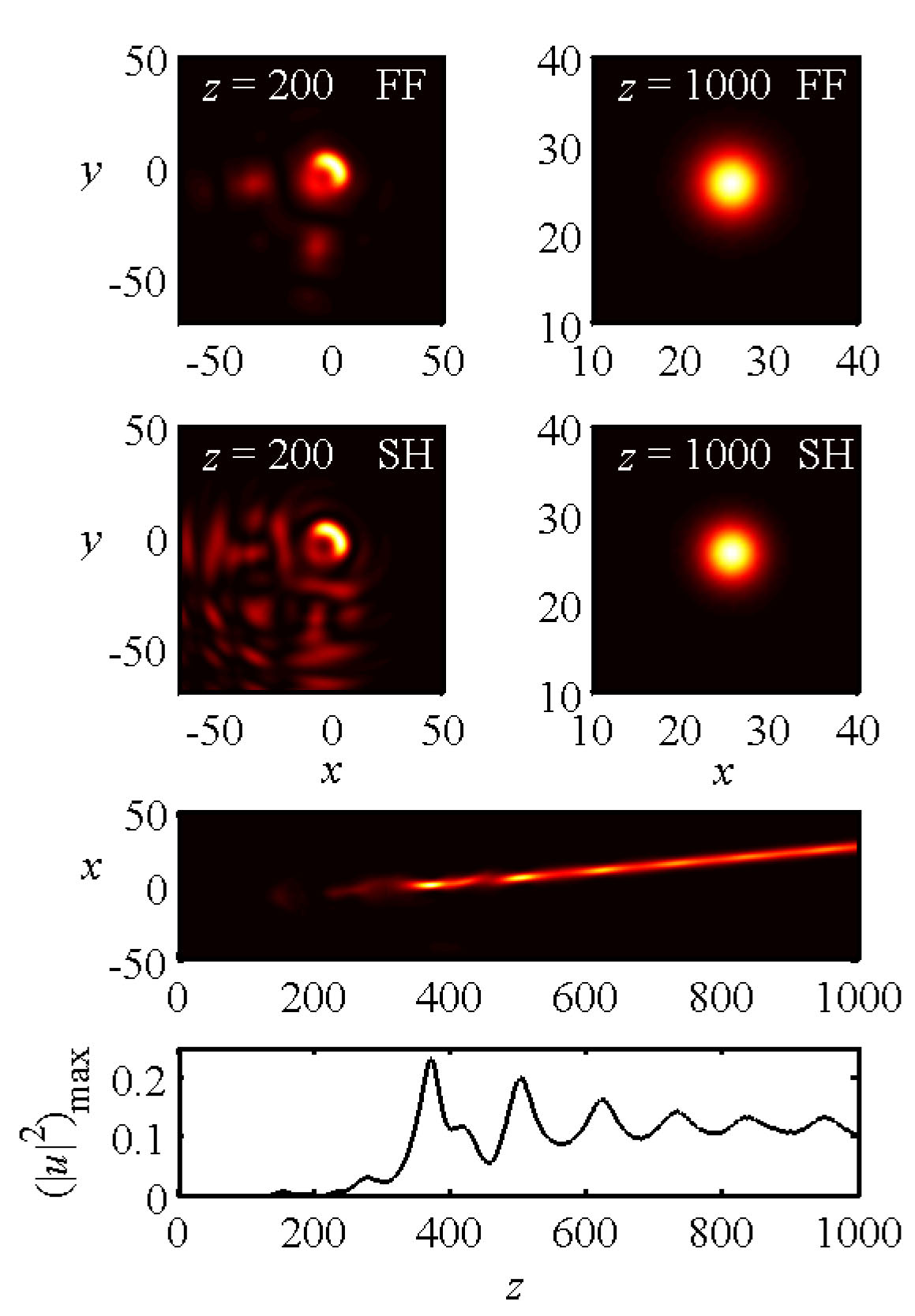}
\caption{A typical example of the generation of a single 2D soliton by input
(\protect\ref{initial}), with $W_{0}=0.3481,$ $\protect\alpha =0.0857$, $A=10
$, and $q=0$ in Eq. (\protect\ref{w}). The first and second rows display,
respectively, distributions of the power density of the FF and SH\
components, $u$ and $w$, in the transverse plane at the early stage of the
evolution ($z=200$), and at $z=1000$, when the well-defined soliton is
observed. The third and fourth rows: The evolution of the FF power-density
profile in the projection onto plane $\left( x,z\right) $, and the peak
power of the FF components vs. $z$, which illustrate the self-trapping of
the soliton.}
\label{Fig2}
\end{figure}

\begin{figure}[tb]
\centering{\includegraphics[width=3in]{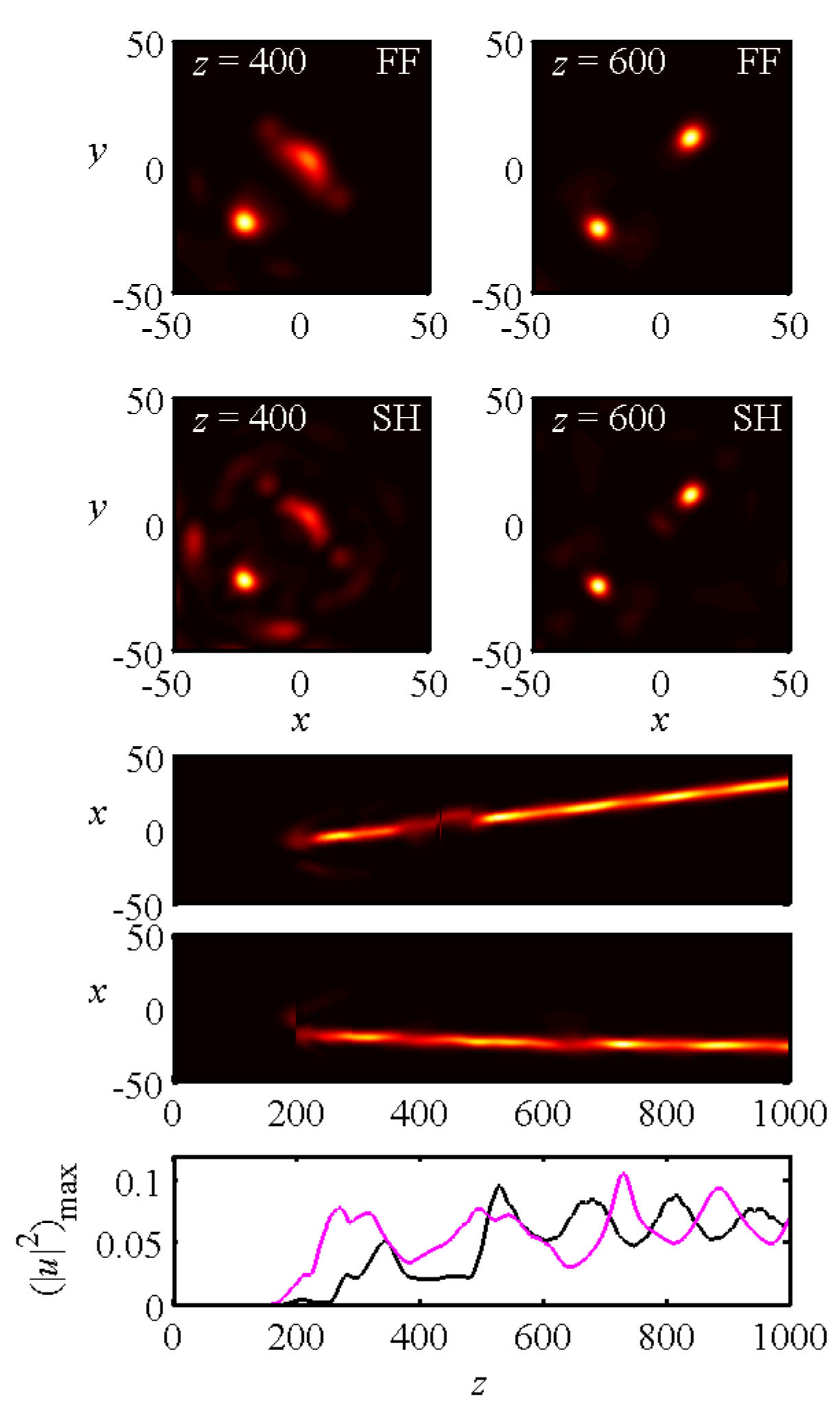}}
\caption{The same as in Fig. \protect\ref{Fig2}, but for a typical example
of the generation of two solitons, at $W_{0}=0.2984$, $\protect\alpha %
=0.05701$, $A=10$, and $q=0$. The third and fourth rows show, severally, the
trajectories of the solitons. In the bottom row, magenta and black lines
show the evolution of the peak powers of the FF component in the left and
right solitons, respectively. }
\label{Fig3}
\end{figure}

Typical examples of the generation of a single soliton and two solitons are
displayed in Figs. \ref{Fig2} and \ref{Fig3}, respectively. In particular,
the single soliton emerging in Fig. \ref{Fig2} traps the share of the
initial total power (\ref{P}) which is $0.251$, the rest being spent on the
generation of small-amplitude radiation (see also Fig. \ref{Fig5} below).
For two solitons in Fig. \ref{Fig3}, respective shares of the initial total
power are $0.200$ and $0.376$. The figures display self-trapping of the
originally diffuse fields, via transient oscillations, into quasi-stationary
2D solitons. Trajectories of the emerging solitons are\ clearly seen too.
Note that, when the single soliton is generated, it moves in the same
direction in which the self-bending AW, initiated by input (\ref{initial}),
would move as the eigenmode of the linear SH\ equation. On the other hand,
in Fig. \ref{Fig3} two solitons move in opposite directions, although the
self-bending direction determined by the AW input remains dominant. It was
checked that the simulations keep zero value of total momentum (\ref{M}),
in accordance with the fact that it is zero for input (\ref{initial}).
The momentum of the moving solitons is compensated by recoil absorbed by the
small-amplitude radiation.

As mentioned above, it the case of $q=0$ in Eq. (\ref{w}), the generation of
three solitons by the AW input (\ref{initial}) requires very small values of
$\alpha $ and, accordingly, a very broad simulation domain. The creation of
multiple solitons is facilitated by taking $q=+1$, as shown in Fig. \ref%
{Fig4}, where an additional weak soliton appears between two stronger ones.
In this case, two emerging solitons move in the direction determined by the
AW input, while the third (left) soliton moves backwards.

\begin{figure}[tb]
\centering{\includegraphics[width=2.8in]{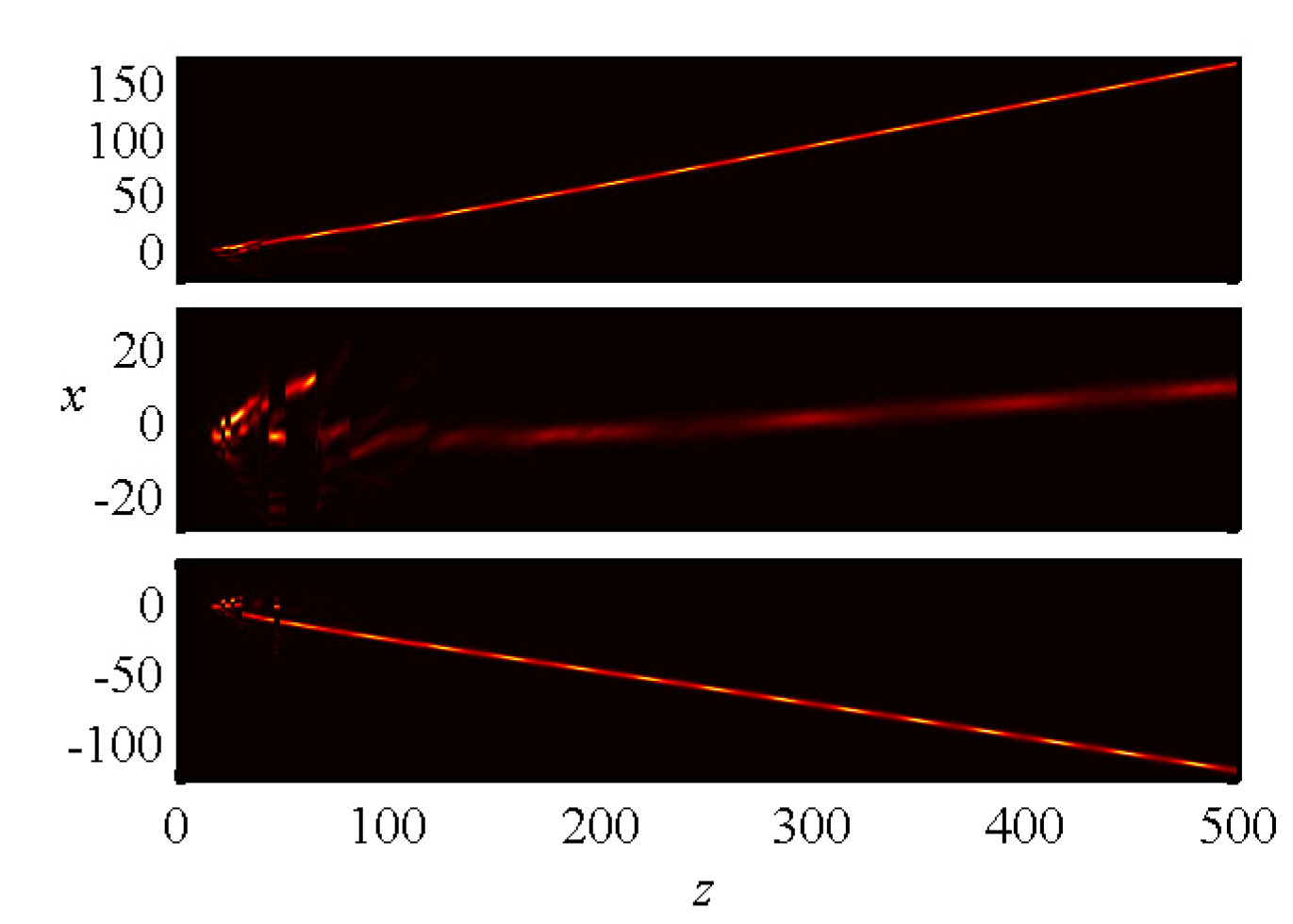}}
\caption{Trajectories of three solitons generated with $W_{0}=3.183,$ $%
\protect\alpha =0.1714$, $A=10$, and $q=+1$ in Eq. (\protect\ref{w}).}
\label{Fig4}
\end{figure}

Unlike the 1D setting considered in Ref. \cite{we}, where relatively strong
radiation \textquotedblleft jets" were generated, in addition to solitons,
making the overall picture rather messy, in the 2D system the radiation
field does not form conspicuous features, allowing the solitons to emerge in
a clean form, as seen in Fig. \ref{Fig5}, where rhombus data points show the
relative intensity of the FF radiation field outside of the soliton's cores.
The result is less clean in the case when two solitons are created, as the
radiation field has a higher amplitude between them.

An essential characteristic of the emerging set of solitons is the share of
total power (\ref{P}) carried away by each one. For the cases when one or
two solitons are generated, the corresponding results are summarized in Fig. %
\ref{Fig5} (in the region shown in the figure, three solitons do not emerge,
but results for the three-soliton set created at $\alpha =0.05$ are
mentioned above, with each soliton's power share close to $0.12$).
Naturally, the shares decrease with the increase of truncation parameter $A$
in input (\ref{initial}), as the total power grows, roughly, $\sim A$, as
per Eq. (\ref{Power}), while parameters of the emerging solitons weakly
depend on $A$. The creation of the solitons is also characterized by the
conversion efficiency, quantified in Fig. \ref{Fig5} by the share of the
total power which is kept by the SH component of the wave field.
\begin{figure}[tb]
\centering{\includegraphics[width=1.9in]{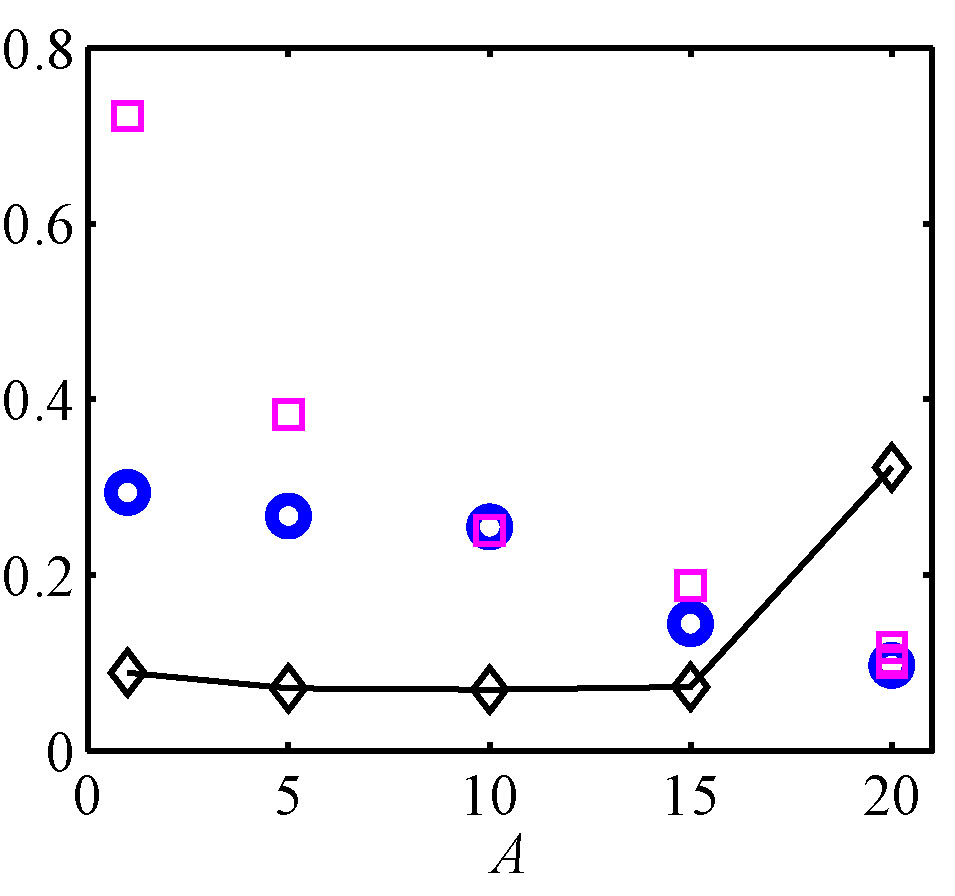}}
\caption{The share of total power (\protect\ref{P}), trapped in the emerging
solitons generated by Eqs. (\protect\ref{u}) and (\protect\ref{w}) with $q=0$%
, is shown by squares as a function of truncation parameter $A$, while $%
\protect\alpha $ is slightly varied to maintain constant FWHM $=20$ of input
(\protect\ref{initial}). The double square at $A=20$ labels the case when
two solitons are created. The cleanness of the emerging solitons is
quantified by rhombuses, which show the relative squared amplitude of the
radiative component of the FF field, $(|u(x,y)|^{2})_{\max }^{\mathrm{(rad)}%
}/(|u(x,y)|^{2})_{\max }^{\mathrm{(sol)}}$), at $z=1000$. Circles show the
share of the total power (\protect\ref{P}) kept by the SH component.}
\label{Fig5}
\end{figure}

\begin{figure}[tb]
\centering\includegraphics[width=3in]{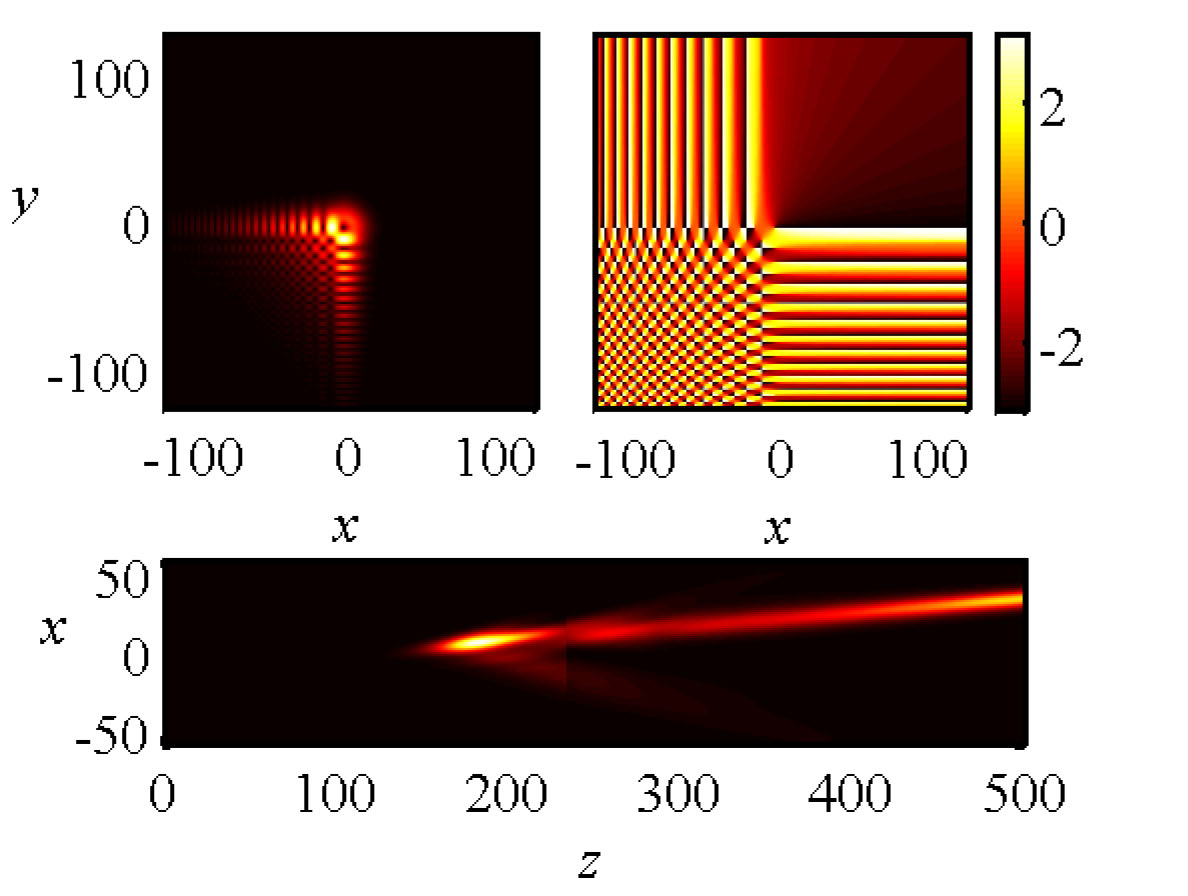}
\caption{Left and right top panels show the amplitude, $|w(x,y)|$, and phase
distributions for the input in the form of the Airy vortex, given by Eq. (%
\protect\ref{vortex}) with $W_{0}=2.6452,$ $\protect\alpha =0.171$, $A=5$.
The bottom panel displays the trajectory of the single fundamental soliton
into which the vortex is converted.}
\label{Fig6}
\end{figure}

We also ran systematic simulations for the transformation of the Airy
vortex, generated by input
\begin{equation}
w\left( x,y,z=0\right) =\left( \partial /\partial x+i\partial /\partial
y\right) \mathrm{AA}\left( x,y\right) ,  \label{vortex}
\end{equation}%
where $\mathrm{AA}\left( x,y\right) $ is the truncated factorized AW in Eq. (%
\ref{initial}). Angular momentum (\ref{Omega}) of this waveform is $\Omega
=4\int \int \left[ \left( \partial \mathrm{AA}/\partial x\right) ^{2}+\left(
\partial \mathrm{AA}/\partial y\right) ^{2}\right] dxdy$. In the framework
of the SH\ linear equation, this input gives rise to a solution produced by
the action of operator $\left( \partial /\partial x+i\partial /\partial
y\right) $ onto the AW.

A typical example of the vortex input and its evolution is displayed in Fig. %
\ref{Fig6}. The mode quickly loses its vorticity and transforms into a
single fundamental soliton, moving in the positive $x$ direction. In the
case shown in Fig. \ref{Fig6}, the total input power (\ref{Power}) is $12.561
$, the emerging soliton carrying away its share $\approx 0.22$. The rest of
the power is scattered in the form of small-amplitude radiation, which also
absorbs the entire angular momentum of the initial vortex.

In conclusion, we have explored the scenario of the downconversion of AWs
(Airy waves) in the 2D $\chi ^{(2)}$medium, which is initiated by random
perturbations amplified by the parametric instability of the AWs and Airy
vortices in the SH component. The spontaneous downconversion generates one
or more stable fundamental solitons in the \textquotedblleft clean" form,
unlike the 1D setting, where the picture is complicated by radiation jets.
A challenging possibility is to extend the analysis
to 3D, where the $\chi ^{(2)}$ nonlinearity supports stable spatiotemporal
solitons \cite{KanRub}-\cite{review}, that may be generated by the 3D
instability of quasi-2D solitons \cite{Wise,DiT,review}.

\section*{Funding Information}

Thailand Research Fund (TRF) (RSA5780061).


\begin{thebibliography}{99}
\bibitem{Berry} M. V. Berry and N. L. Balazs, Am. J. Phys. \textbf{47}, 264
(1979).

\bibitem{Berry-tsunami} M. V. Berry, New J. Phys. \textbf{7}, 129 (2005).

\bibitem{Chr1} G. A. Siviloglou and D. N. Christodoulides, Opt. Lett.
\textbf{99}, 213901 (2007).

\bibitem{Chr2} G. A. Siviloglou, J. Broky, A. Dogariu, and D. N.
Christodoulides, Phys. Rev. Lett. \textbf{99}, 213901 (2007).

\bibitem{Chr3} P. Polynkin, M. Kolesik, J. V. Moloney, G. A. Siviloglou, and
D. N. Christodoulides, Science \textbf{324}, 229 (2009).

\bibitem{Denz} P. Rose, F. Diebel, M. Boguslawski, and C. Denz, Appl. Phys.
Lett. \textbf{102}, 101101 (2013).

\bibitem{inversion} R. Driben, Y. Hu, Z. Chen, B. A. Malomed, and R.
Morandotti, Opt. Lett. \textbf{38}, 2499 (2013).

\bibitem{Efremidis} N. K. Efremidis, Phys. Rev. A \textbf{98}, 023841 (2014).

\bibitem{Milivoj} Y. Zhang, M. R. Beli\'{c}, L. Zhang, W. P. Zhong, D. Y.
Zhu, R. M. Wang, and Y. P. Zhang, Opt. Exp. \textbf{23}, 10467 (2015).

\bibitem{plasm1} A. Minovich, A. E. Klein, N. Janunts, T. Pertsch, D. N.
Neshev, and Y. S. Kivshar, Phys. Rev. Lett. \textbf{107}, 116802 (2011).

\bibitem{plasm2} L. Li, T. Li, S. M. Wang, C. Zhang, and S. N. Zhu, Phys.
Rev. Lett. \textbf{107}, 126804 (2011).

\bibitem{plasm3} I. Epstein and A. Arie, ``Arbitrary Bending Plasmonic Light
Waves", Phys. Rev. Lett. \textbf{112}, 023903 (2014).

\bibitem{el} N. Voloch-Bloch, Y. Lereah, Y. Lilach, A. Gover, and A. Arie,
Nature \textbf{494}, 331 (2013).


\bibitem{discharge} M. Clerici, Y. Hu, P. Lassonde, C. Mili\'{a}n, A.
Couairon, D. N. Christodoulides, Z. Chen, L. Razzari, F. Vidal, F. L\'{e}gar%
\'{e}, D. Faccio, R. Morandotti, Sci. Adv. \textbf{1}, e140011 (2015).


\bibitem{1Dand2Dexperiment0} D. G. Papazoglou, S. Suntsov, D. Abdollahpour,
and S. Tsortsakis, Phys. Rev. A \textbf{81}, 061807 (2010).

\bibitem{1Dand2Dexperiment} Z. Y. Ye, S. Liu, C. B. Lou, P. Zhang, Y. Hu, D.
H. Song, J. L. Zhao, and Z. G. Chen, Opt. Lett. 36, 3230 (2011).

\bibitem{ChrSegev-nonlin} I. Kaminer, M. Segev, and D. N. Christodoulides,
Phys. Rev. Lett. \textbf{106}, 213903 (2011).

\bibitem{Marom} Y. Fattal, A. Rudnick, and D. M. Marom, Opt. Exp. \textbf{19}%
, 17298 (2011).


\bibitem{Morandotti} Y. Hu, Z. Sun, D. Bongiovanni, D. Song, C. Lou, J. Xu,
Z. Chen, and R. Morandotti, Opt. Lett. \textbf{37}, 3201 (2012).


\bibitem{Radik-VVK} R. Driben, V. V. Konotop, and T. Meier, Opt. Lett.
\textbf{39}, 5523 (2014).

\bibitem{Porras} C. Ruiz-Jim\'{e}nez, K. Z. N\'{o}brega, and M. A. Porras,
Opt. Exp. \textbf{23}, 8918 (2015).

\bibitem{Radik} R. Driben and T. Meier, Opt. Lett. \textbf{39}, 5539,
(2014).

\bibitem{Ady-3wave} T. Ellenbogen, N. Voloch-Bloch, A. Ganany-Padowicz, and
A. Arie, Nature Phot. \textbf{3}, 395 (2009).


\bibitem{Ady-Moti} I. Dolev, I. Kaminer, A. Shapira, M. Segev, and A. Arie,
Phys. Rev. Lett. \textbf{108}, 113803 (2012).

\bibitem{2DexperimentChi2} I. Dolev and A. Arie, Appl. Phys. Lett. \textbf{97%
}, 171102 (2010).

\bibitem{rev1} G. I. Stegeman, D. J. Hagan, and L. Torner, Opt. Quant.
Elect. \textbf{28}, 1691 (1996).

\bibitem{rev2} C. Etrich, F. Lederer, B. A. Malomed, T. Peschel, and U.
Peschel, Prog. Opt. \textbf{41}, 483 (2000).

\bibitem{rev3} A. V. Buryak, P. Di Trapani, D. V. Skryabin, and S. Trillo,
Phys. Rep. \textbf{370}, 63 (2002).

\bibitem{we} T. Mayteevarunyoo and B. A. Malomed, Opt. Lett. \textbf{40},
4947 (2015).

\bibitem{down1D}
G. Leo and G. Assanto, 
Opt. Lett.  {\bf 22}, 1391
(1997).

\bibitem{down2D} M. T. G. Canva, R. A. Fuerst, S. Baboiu, and G. I.
Stegeman, and G. Assanto, 
Opt. Lett. \textbf{22}, 1683
(1997).

\bibitem{2Dchi2solitons} W. E. Torruellas, Z. Wang, D. J. Hagan, E. W. Van
Stryland, G. I. Stegeman, L. Torner, and C. R. Menyuk, Phys. Rev. Lett.
\textbf{74}, 5036 (1995).

\bibitem{vort1} W. J. Firth and D. V. Skryabin, Phys. Rev. Lett. \textbf{79}%
, 2450 (1997).

\bibitem{vort2} L. Torner and D. V. Petrov, Electron. Lett. \textbf{33}, 608
(1997).

\bibitem{vort3} D. V. Petrov, L. Torner, J. Martorell, R. Vilaseca, J.P.
Torres and C. Cojocaru, Opt. Lett. \textbf{23}, 1444 (1998).

\bibitem{Spain} J. J. Gar\'{c}ia-Ripoll, V. V. Konotop, B. Malomed, and V.
M. P\'{e}rez-Garc\'{\i}a, Mathematics and Computers in Simulation \textbf{62}%
, 21 (2003).

\bibitem{Fibich} G. Fibich, The Nonlinear Schr\"{o}dinger Equation: Singular
Solutions and Optical Collapse (Springer: Cham, 2015).

\bibitem{KanRub} A. A. Kanashov and A. M. Rubenchik, Physica D \textbf{4},
122 (1980).

\bibitem{HH} B. A. Malomed, P. Drummond, H. He, A. Berntson, D. Anderson,
and M. Lisak, Phys. Rev. E \textbf{56}, 4725 (1997).

\bibitem{review} B. A. Malomed, D. Mihalache, F. Wise, and L. Torner, J.
Optics B: Quant. Semicl. Opt. \textbf{7}, R53 (2005).

\bibitem{Wise} X. Liu, K. Beckwitt, K, and F. Wise, Phys. Rev. Lett. \textbf{%
85}, 1871 (2000).

\bibitem{DiT} S. Minardi, J. Yu, G. Blasi, A. Varanavi\v{c}ius, G. Valiulis,
A. Ber\v{z}anskis, A. Piskarskas, and P. Di Trapani, Phys. Rev. Lett.
\textbf{91}, 123901 (2003).
\end{thebibliography}
\end{document}